\documentclass[twocolumn,showpacs,preprintnumbers,amsmath,amssymb,superscriptaddress]{revtex4}

\usepackage{graphicx}
\begin{document}

\title{Multiple-type solutions for multipole interface solitons in thermal nonlinear media}

\author{Xuekai Ma, Zhenjun Yang, Daquan Lu, Wei Hu}

\email[Corresponding author's email address: ]{huwei@scnu.edu.cn}

\address{Laboratory of Photonic Information Technology, South China
Normal University, Guangzhou 510631, P. R. China}

\date{\today}

\begin{abstract}
We address the existence of multipole interface solitons in
one-dimensional thermal nonlinear media with a step in the linear
refractive index at the sample center. It is found that there exist
two types of solutions for tripole and quadrupole interface
solitons. The two types of interface solitons  have different
profiles, beam widths, mass centers, and stability regions. For a
given propagation constant, only one type of interface soliton is
proved to be stable, while the other type can also survive over a
long distance. In addition, three types of solutions for fifth-order
interface solitons are found.
\end{abstract}


\maketitle

\section{introduction}
Nonlocality of a nonlinear response is a property exhibited in many
nonlinear optical media. Nonlocal solitons have been found in
nematic liquid
crystals~\cite{Peccianti-2002-OL,Rasmussen-2005-PRE,Hu-2006-APL,Conti-2003-PRL,Conti-2004-PRL}
and lead
glasses~\cite{Rotschild-2005-PRL,Rotschild-2006-NP,Rotschild-2006-OL,Skupin-2006-PRE}
theoretically and experimentally. They present some novel
properties, for instance, the large phase shift~\cite{Guo-2004-PRE},
self-induced fractional Fourier transform~\cite{Lu-2008-PRA},
attraction between two dark solitons~\cite{Nikolov-2004-OL}, etc.
Recently, various types of nonlocal surface
solitons~\cite{Alfassi-2007-PRL,Kartashov-2009-OL,Ye-2008-PRA,Alfassi-2009-PRA,Kartashov-2006-OL,Kartashov-2007-OL,Ma-2011-PRA},
for example, multipole surface
solitons~\cite{Kartashov-2009-OL,Ye-2008-PRA}, vortex surface
solitons~\cite{Ye-2008-PRA}, and incoherent surface
solitons~\cite{Alfassi-2009-PRA}, have been found at the interface
between a nonlinear medium and a linear
medium~\cite{Alfassi-2007-PRL,Kartashov-2009-OL,Ye-2008-PRA,Alfassi-2009-PRA,Kartashov-2007-OL}
or between two nonlinear media~\cite{Kartashov-2006-OL,Ma-2011-PRA}.
The surface solitons propagating at an interface formed by a
nonlinear medium and a linear medium are found to be stable only
when their peaks are less than three~\cite{Kartashov-2009-OL}.
Surface solitons propagating at an interface formed by two nonlinear
media are also found~\cite{Kartashov-2006-OL}, but dipole surface
solitons in such media can exist only when optical lattices exist.

Nonlocal multipole solitons are studied in nematic liquid
crystals~\cite{Xu-2005-OL} and lead
glass~\cite{Dong-2010-PRA,Kartashov-2009-OL} for both bulk solitons
and surface solitons. In nonlocal bulk media, multipole solitons are
symmetric, and they are stable if they contain fewer than five
peaks~\cite{Xu-2005-OL,Dong-2010-PRA}. For surface multipole
solitons, the profiles are asymmetric because of the existence of
boundary conditions, and they are stable when the number of
intensity peaks is less than three~\cite{Kartashov-2009-OL}. By
comparison, we can qualitatively consider surface solitons as half
of their corresponding solitons in bulk media~\cite{Ma-2011-PRA1}.
For example, surface fundamental solitons can be regarded as half of
dipole solitons in bulk media. However, for both bulk solitons and
surface solitons in nonlocal media, only one type of solution exists
in any case.

In Ref. \cite{Ma-2011-PRA}, we address the existence of the
fundamental and dipole interface solitons propagating at the
interface between two thermal nonlinear media with different linear
refractive indices. Fundamental interface solitons are found to
always be stable, and the stability of dipole interface solitons
depends on the difference in linear refractive index. The boundary
force effect~\cite{Alfassi-2007-OL} plays an important role in the
stability of dipole interface solitons. It is found that the mass
center of the fundamental and dipole interface solitons moves to the
part with higher linear refractive index as the index difference
between two media increases~\cite{Ma-2011-PRA}.

In this paper, we study multipole interface solitons in thermal
nonlinear media. It is found that there exist two types of tripole
and quadrupole interface solitons and three types of fifth-order
interface solitons, respectively. The phenomenon of two (or three)
soliton types is not found in other surface solitons or in bulk
solitons.

\section{model of interface solitons}
We consider a (1+1)dimensional thermal sample occupying the region
$-L\leq x\leq L$. The sample is separated into two parts by the
interface at $x=0$. All parameters for the two parts are same except
the linear refractive indices. The propagation of a TE polarized
laser beam is governed by the dimensionless nonlinear
Schr\"{o}dinger equation,\\
(i) on the left, i.e., $-L\leq x\leq0$,
\begin{equation}\label{1}
  i\frac{\partial q}{\partial
z}+\frac{1}{2}\frac{\partial^2q}{\partial x^2}+nq=0,\,\,\,\,\,\,\,
  \frac{\partial^2n}{\partial x^2}=-|q|^2,
\end{equation}
(ii) on the right, i.e., $0\leq x\leq L$,
\begin{equation}\label{2}
  i\frac{\partial q}{\partial
z}+\frac{1}{2}\frac{\partial^2q}{\partial x^2}+nq-n_dq=0,\,\,\,\,
  \frac{\partial^2n}{\partial x^2}=-|q|^2,
\end{equation}
where $x$ and $z$ stand for the normalized transverse and
longitudinal coordinates, $q$ is the complex amplitude of the
optical field, $n$ is the nonlinear refractive index, and $n_d>0$ is
the difference in linear refractive index between two media. Two
boundaries ($x=\pm L$) and the interface ($x=0$) are thermally
conductive. Boundary conditions can be described by $q(\pm L)=0$ and
$n(\pm L)=0$, and the continuity conditions at the interface are
$q(-0)=q(+0)$ and $n(-0)=n(+0)$. Thus, both $q$ and $n$ are
continuous at the interface.

We search for soliton solutions for Eqs.~(\ref{1}) and (\ref{2})
numerically in the form $q(x,z)=w(x)\exp(ibz)$, where $w(x)$ is a
real function and $b$ is the propagation constant. The details of
the model and results for fundamental and dipole interface solitons
can be found in Ref.~\cite{Ma-2011-PRA}. To elucidate the stability
of interface solitons, we search for the perturbed solutions for
Eqs. (\ref{1}) and (\ref{2}) in the form $q=(w+u+iv)\exp(ibz)$,
where $u(x,z)$ and $v(x,z)$ are the real and imaginary parts of the
small perturbations. The perturbation can grow with a complex rate
$\sigma$ upon propagation. Substituting the perturbed soliton
solution into Eqs. (\ref{1}) and (\ref{2}), one can get the linear
eigenvalue problem,
\begin{equation}\label{3}
 \left.\begin{aligned}
  \sigma u&=-\frac{1}{2}\frac{d^2v}{dx^2}+bv-nv,  \\
  \sigma v&=\frac{1}{2}\frac{d^2u}{dx^2}-bu+nu+w\Delta n,  \\
\end{aligned}\right\}
\,\,\,\, \text{($-L\leq x\leq0$),}
\end{equation}
and
\begin{equation}\label{4}
 \left.\begin{aligned}
  \sigma u&=-\frac{1}{2}\frac{d^2v}{dx^2}+bv-nv+n_dv,  \\
  \sigma v&=\frac{1}{2}\frac{d^2u}{dx^2}-bu+nu-n_du+w\Delta n,  \\
\end{aligned}\right\}
\,\,\,\, \text{($0\leq x\leq L$),}
\end{equation}
where $\Delta n=-2\int_{-L}^{L}G(x,x')w(x')u(x')dx'$ is the
refractive index perturbation, the response function
$G(x,x')=(x+L)(x'-L)/(2L)$ for $x\leq x'$, and
$G(x,x')=(x'+L)(x-L)/(2L)$ for $x\geq x'$~\cite{Dong-2010-PRA}.

\section{multipole interface solitons}

\begin{figure}[htb]
\centerline{\includegraphics[width=8.5cm]{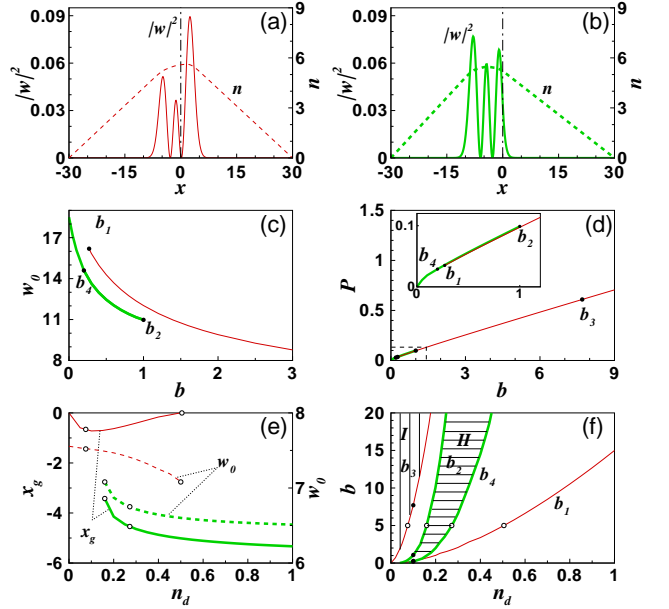}} \caption{(Color
online) (a) and (b) Profiles for two types of tripole interface
solitons at $n_d=0.4$ and $b=5$. Dashed-dotted lines stand for the
interface. (c) Beam width versus propagation constant at $n_d=0.1$.
(d) Soliton power versus  propagation constant at $n_d=0.1$, where
the inset is enlarged for small values of $b$. (e) Mass center and
beam width versus the index difference at $b=5$.
 (f) Regions of existence and stability of tripoles in the $b-n_d$ plane, where shadow areas show the stable regions.
Points correspond to those cases in (c) and (d), and circles
correspond to those cases in (e). Red (thin) and green (thick) lines
stand for type-I and type-II tripoles, respectively, in (a)-(f).
}\label{f1}
\end{figure}

The results of tripole interface solitons are shown in
Fig.~\ref{f1}. The most interesting feature of tripole interface
solitons is that there exist two different types of solutions for
some given values of $n_d$ and $b$.
For example, when $n_d=0.4$ and $b=5$, two solutions are shown in
Figs.~\ref{f1}(a) and \ref{f1}(b), respectively. For the first type
of solution (named type I in this paper) as shown in
Fig.~\ref{f1}(a), two intensity peaks are located in the left part
of the sample with a higher index, and one resides in the right part
with a lower index. The right peak is much higher than the left two
peaks because the peak of nonlinear refractive index $n$ is located
at the right side of the interface. For the type-II solution shown
in Fig.~\ref{f1}(b), almost all three peaks are located in the left
part of the sample, and the left peak is the highest. Due to the
difference in the profile, the beam widths (in root-mean-square
definition) for the two types of tripoles are different as shown in
Fig. \ref{f1}(c), and they decrease monotonically with increasing
$b$. However, their powers, defined as $P=\int_{-L}^L|q|^2dx$ and
increasing monotonically with increasing $b$, are approximately
equal (relative difference is smaller than 2\%) as shown in
Fig.~\ref{f1}(d).

It is known that the interface soliton in our model will reduce to
the soliton in bulk media when $n_d=0$. As $n_d$ increases, the
tripole interface soliton becomes asymmetric and shifts to the left
part with a higher index. From Fig.~\ref{f1}(e), only the type-I
tripole exists for a small value of the index difference, and its
mass center (defined as $x_g=\int_{-L}^L x|q|^2dx/P$) and center
peak move into the left part as $n_d$ increases. When $n_d$
overtakes a certain value at a fixed $b$, for example, $n_d \geq
0.16$ at $b=5$, there exist two types of tripoles [Fig.
\ref{f1}(f)]. It is interesting to note that the right peak of the
type-I tripole always remains in the right part, and its mass center
moves back to the interface as $n_d$ increases sequentially [Figs.
\ref{f1}(a) and \ref{f1}(e)]. For the type-II tripole, its right
peak moves into the left part of the sample, and its mass center
shifts toward the left and approaches a fixed value when $n_d$ is
large enough, for example, $x_g\rightarrow-6.5$ at $b=5$ as shown in
Fig.~\ref{f1}(e). The asymptotic feature is induced by the boundary
force effect~\cite{Alfassi-2007-OL}, and it is similar to that of
fundamental and dipole interface solitons~\cite{Ma-2011-PRA}. The
mass center of the type-II tripole is proportional to its beam
width, whereas, the change of mass center for the type-I tripole is
irrelative with its beam width [Fig. \ref{f1}(e)].

The existence regions of tripole interface solitons are found
numerically as $b\geq b_1$ (type I) or $b\leq b_2$ (type II) for a
given index difference, where $b_1$ and $b_2$ are critical
propagation constants and are given in Figs. \ref{f1}(c),
\ref{f1}(d), and \ref{f1}(f). In the overlay region ($b_1\leq b \leq
b_2$), which increases as $n_d$ increases, shown in Fig.
\ref{f1}(f), two types of tripoles can exist simultaneously for a
fixed $n_d$. For a given propagation constant, there also exists a
region of $n_d$ in which the two types of tripoles exist
simultaneously. If $b=5$,  for example, the type-I tripoles exist in
$n_d \leq 0.51$ and the type-II tripoles exist in $n_d \geq 0.16$,
then the overlay region is  $0.16 \leq n_d \leq 0.51$ [see the
circle symbols in Figs. \ref{f1}(e) and \ref{f1}(f)].


\begin{figure}[htb]
\centerline{\includegraphics[width=8.5cm]{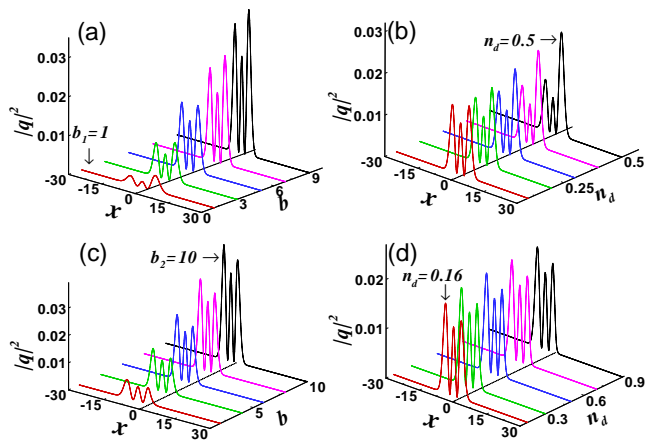}} \caption{(Color
online) Soliton intensity profiles vary (a) with $b$ at $n_d=0.2$
and (b) with $n_d$ at $b=5$ for type-I solutions of tripole
interface solitons. Soliton intensity profiles vary (c) with $b$ at
$n_d=0.2$ and (d) with $n_d$ at $b=5$ for type-II solutions of
tripole interface solitons.}\label{f5}
\end{figure}

Figure \ref{f5} shows the intensity peaks of tripole solitons vary
with $b$ and $n_d$ for both type-I and type-II solutions. From Figs.
\ref{f5}(a) and \ref{f5}(c), we can see that the soliton power
increases with an increase in the constant propagation at a fixed
$n_d$, which has been shown in Fig. \ref{f1}(d). For a fixed $n_d$
(for example, $n_d=0.2$), there exist critical propagation constants
for both type-I and type-II solutions as shown in Fig. \ref{f1}(f).
In Fig. \ref{f5}(a), the critical propagation constant ($b_1=1$) is
marked, and the type-I solution of the solitons disappears when
$b<b_1$. For type-II solutions, when $b>b_2=10$, marked in Fig.
\ref{f5}(c), one cannot find this kind of solution anymore. If we
fix the propagation constant, there exists critical $n_d$ for both
type-I and type-II solutions as shown in Fig. \ref{f1}(f), which
also can be explained by Figs. \ref{f5}(b) and \ref{f5}(d). Figures
\ref{f5}(b) and \ref{f5}(d) present the change in intensity peaks
versus $n_d$. For $n_d>0.5$, the type-I solutions disappear [Fig.
\ref{f5}(b)], while  one cannot find the type-II solutions for
$n_d<0.16$ [Fig. \ref{f5}(d)]. It is noted that, for type-I
solutions, the changes in the intensity peaks are obvious [Fig.
\ref{f5}(b)], while the intensity peaks of type-II solutions change
little [Fig. \ref{f5}(d)].

The stability regions of tripole interface solitons are found as
$b\geq b_3$ (type I) and $b_4\leq b\leq b_2$ (type II) for a given
$n_d$, where $b_3$ and $b_4$ are shown in Fig.~\ref{f1}(f). The
letters I (shadow region with vertical lines) and II (shadow region
with horizontal lines) in Fig.~\ref{f1}(f) indicate the stable zones
in the $b-n_d$ plane. It is obvious that the two types of tripoles
can not be stable simultaneously for a given propagation constant.
In the region $b_4\leq b \leq b_2$, the type-II tripole is stable,
but the type-I tripole is not.

For comparison, tripole solitons in bulk thermal nonlinear media are
stable~\cite{Rotschild-2006-OL,Dong-2010-PRA}, but tripole surface
solitons at the interface between a thermal nonlinear medium and a
linear medium are unstable~\cite{Kartashov-2009-OL}. From
Fig.~\ref{f1}(f), $b_1=0$ for type-I solutions when $n_d=0.06$ and
$b_2=0$ for type-II solutions when $n_d=0.05$. Only the type-I
tripoles exist for $n_d<0.05$, and they are stable almost in their
whole domain since $b_3$ approaches zero when $n_d$ approaches zero.
For a very large $n_d$, it is noted that both types of tripole
interface solitons have their stability regions, unlike their
counterparts in surface solitons. For the type-II tripoles, almost
all the energy resides in the higher-index part, which is similar to
the fundamental and dipole interface solitons~\cite{Ma-2011-PRA} and
surface
solitons~\cite{Alfassi-2007-PRL,Kartashov-2009-OL,Ye-2008-PRA}. For
the type-I tripoles, the right peak still resides in the lower-index
medium even for a large $n_d$ [Fig. \ref{f5}(b)]. This unusual
feature reasonably cannot be explained only by the boundary force
effect, since the type-I tripole with smaller $|x_g|$ (closer to the
center of the sample) should get the less equivalent force from the
boundaries than the type-II tripole.


\begin{figure}[htb]
\centerline{\includegraphics[width=8.5cm]{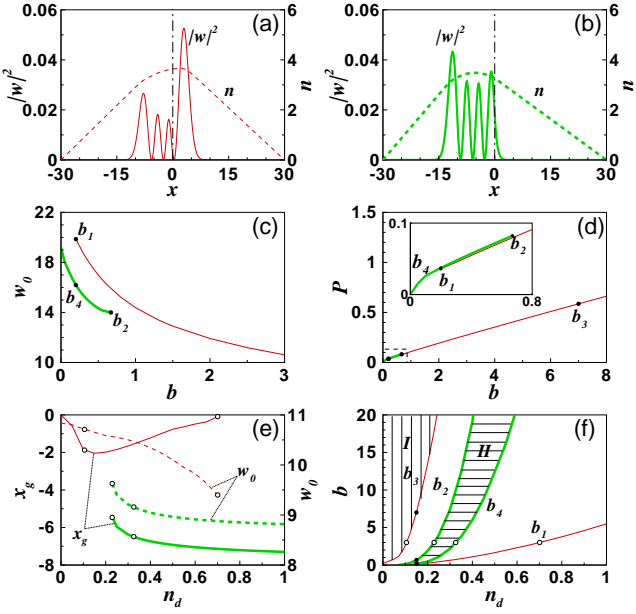}} \caption{(Color
online) (a) and (b) Profiles for two types of quadrupole interface
solitons at $n_d=0.5$ and $b=3$. (c) Beam width versus propagation
constant  at $n_d=0.15$. (d) Soliton power versus propagation
constant  at $n_d=0.15$, where the inset is enlarged for small
values of $b$. (e) Mass center and beam width versus the index
difference at $b=3$. (f) Regions of existence and stability of
quadrupoles in the $b-n_d$ plane. The style of each figure is the
same as that in Fig.\ref{f1}.}\label{f2}
\end{figure}

The results of quadrupole interface solitons are shown in Fig.
\ref{f2}, which are very similar to tripole interface solitons.
There also exist two types of quadrupole interface solitons as shown
in Figs. \ref{f2}(a) and \ref{f2}(b). The type-I quadrupole has
three peaks in the left and one peak in the right, whereas, the
type-II solution has four peaks in the left part. The properties of
the quadrupoles are the same as that of the tripoles, except for the
value of the parameters. It is worthy to note that there exist
stability regions, i.e., $b\geq b_3$ (region I with vertical lines)
for type-I solutions and $b_4\leq b\leq b_2$ (region II with
horizontal lines) for type-II solutions, for very large index
differences. Also, we have observed that there is one peak of the
type-I quadrupole still residing in the lower-index part even for a
very large $n_d$ [Fig. \ref{f2}(a)].

\begin{figure}[htb]
\centerline{\includegraphics[width=8.5cm]{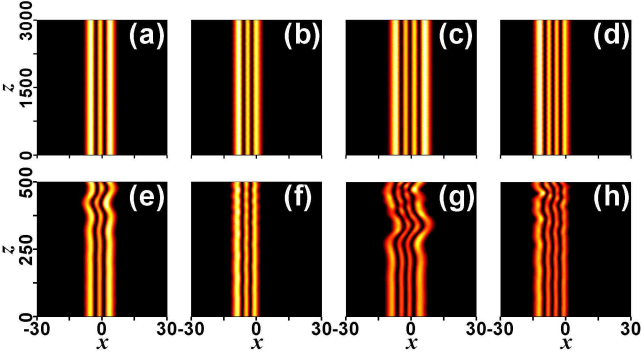}} \caption{(Color
online) Propagations of tripole interface solitons at $b=3$ for (a)
$n_d=0.05$, (b) $n_d=0.15$, (e) $n_d=0.1$, and (f) $n_d=0.25$.
Propagations of quadrupole interface solitons at $b=2$ for (c)
$n_d=0.05$, (d) $n_d=0.25$, (g) $n_d=0.15$, and (h)
$n_d=0.35$.}\label{f3}
\end{figure}

To confirm the results of the linear stability analysis, we simulate
the soliton propagation based on Eqs.~(\ref{1}) and (\ref{2}) with
the input condition $q(x,z=0)=w(x)[1+\rho(x)]$, where $w(x)$ is the
profile of the stationary wave and $\rho(x)$ is a random function
that stands for the input noise with the variance
$\delta_{noise}^2=0.01$. Figure \ref{f3} presents propagations of
tripole and quadrupole interface solitons for both type-I [Figs.
\ref{f3}(a), \ref{f3}(c), \ref{f3}(e), and \ref{f3}(g)] and type-II
[Figs. \ref{f3}(b), \ref{f3}(d), \ref{f3}(f), and \ref{f3}(h)]. As
expected, multipole interface solitons, in their stability regions,
survive over long propagation distances in the presence of the input
noise (the top row). The bottom row in Fig.~\ref{f3} presents
propagations of multipole interface solitons in their instability
regions. They experience oscillatory instability after propagating
over a long distance ($>$ 200 Rayleigh distances). This distance is
long enough to observe the interface solitons in experiments, so one
experimentally can observe the two types of tripole (quadrupole)
interface solitons at the same $n_d$ and $b$.

\begin{figure}[htb]
\centerline{\includegraphics[width=8.5cm]{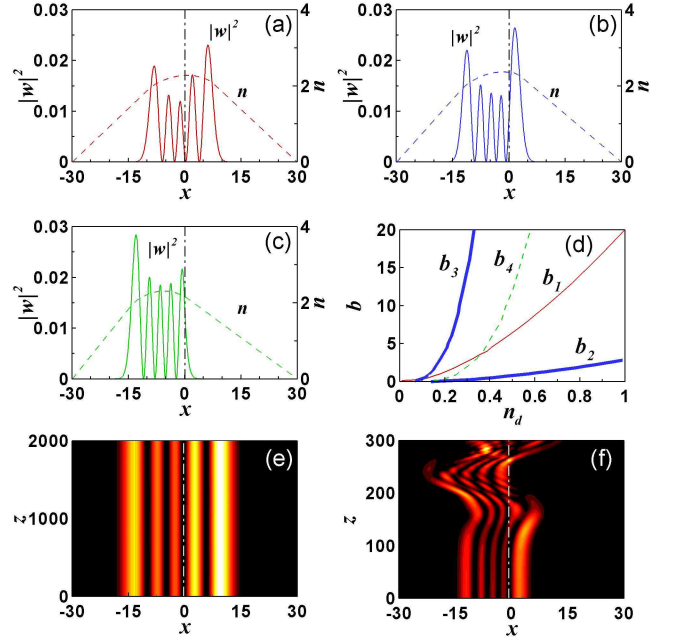}} \caption{(Color
online) Profiles of fifth-order interface solitons at (a) $n_d=0.2$,
(b) $n_d=0.5$, and (c) $n_d=0.5$. For all cases $b=3$. (d) Regions
of the existence of fifth-order interface solitons in the $b-n_d$
plane. Propagations of fifth-order interface solitons at (e)
$n_d=0.129$, $b=0.5$ and (f) $n_d=0.5$, $b=2$.}\label{f4}
\end{figure}

Figure \ref{f4} shows the results of fifth-order interface solitons
at different conditions. There exist three types of fifth-order
interface solitons. If $n_d$ is small, only one solution exists with
three peaks on the left and two peaks on the right (type I) as shown
in Fig. \ref{f4}(a). When $n_d$ increases, this type of solution
disappears, and other two types of solutions exist as shown in Figs.
\ref{f4}(b) and \ref{f4}(c). In Fig. \ref{f4}(b), there are four
peaks on the left and one on the right (type II), whereas, in Fig.
\ref{f4}(c), almost all five peaks reside on the left (type III).
The existence regions of the three types of solutions for
fifth-order interface solitons are shown in Fig. \ref{f4}(d). The
type-I solutions exist in the region $b\geq b_1$ [above the red
(thin) line], and type-II solutions can be found in the region
$b_2\leq b\leq b_3$ [between the two blue (thick) lines]. At the
region $b\leq b_4$ [below the green (dashed) line], one can find the
type-III solutions. It is found that the number of solution types is
different for different values of $n_d$ and $b$ [Fig. \ref{f4}(d)].
For example, when $b< b_2$ or $b>b_3$, only one type of solution
exists. When $n_d>0.36$ and $b_1<b<b_4$, three types of solutions
can exist simultaneously.

It is known that the solitons in bulk thermal media are unstable
when the number of peaks is more than
4~\cite{Dong-2010-PRA,Xu-2005-OL}. However, for the fifth-order
interface solitons, there exists a stability region, although this
region (not given here) is very small. One of the major reasons for
the stable fifth-order interface solitons is the existence of the
interface. For type-I solutions, one can find the stable fifth-order
interface solitons when both $n_d$ and $b$ are small [Fig.
\ref{f4}(e)]. However, there do not exist stable fifth-order
interface solitons for type-II and type-III solutions. Although the
two types of solutions are unstable, they can propagate a long
distance as shown in Fig. \ref{f4}(f). In addition, for the
higher-order (more than five) interface solitons, stability regions
almost do not exist.

\section{conclusion}
To conclude, we have studied the properties  of multipole interface
solitons in thermal nonlinear media. It is found that there exist
two different types of tripole-quadrupole interface solitons, and
three types of  fifth-order interface solitons. When the linear
index difference between two media is small, only the type-I
solutions of the tripole and quadrupole interface solitons exist,
and they are stable almost in their whole domain. When the index
difference is large, two types of tripole and quadrupole interface
solitons can exist stably in different stability regions, unlike
their counterparts in surface solitons. It is unusual that the right
peak of the type-I multipole interface soliton still resides in the
lower-index medium even for a very large $n_d$. In addition, our
model can support stable high-order solitons with more than four
peaks, which is quite different from that in uniform media, although
there only exists a small stability region for the interface
solitons with more than four intensity peaks. The concept mentioned
in this paper can be extended to other nonlinear systems.

\section*{acknowledge}
This research was supported by the National Natural Science
Foundation of China (Grants No. 10804033 and No. 10674050) and the
Specialized Research Fund for the Doctoral Program of Higher
Education (Grant No. 200805740002).

\end{document}